\documentclass[fleqn,12pt]{wlscirep}
\usepackage[utf8]{inputenc}
\usepackage[T1]{fontenc}
\usepackage{times}
\usepackage{graphicx}
\usepackage{color}
\usepackage{ragged2e}
\usepackage{amsmath}
\usepackage{braket}
\usepackage{lineno}
\usepackage{xcolor}
\usepackage{xpatch}
\usepackage{cite}
\usepackage{float}  
\newfloat{extended_fig}{htbp}{loflow} 
\floatname{extended_fig}{Extended Data Figure}  

\title{Revisiting self-seeding mechanism by generating vector ultraviolet N$_2^{+}$ lasing}

\author[1,2,$\ast$]{Jingsong Gao}
\author[1]{Yang Wang}
\author[3]{Haicheng Mei}
\author[3]{Jiahao Dong}
\author[3]{Yi Liu}
\author[1]{Chengyin Wu}
\author[1,$\dagger$]{Hongbing Jiang}
\author[2]{Meng Han}
\author[1,$\ddagger$]{Yunquan Liu}

\affil[1]{State Key Laboratory for Mesoscopic Physics and Frontier Science Center for Nano-Optoelectronic, School of Physics, Peking University, Beijing, 100871, China}

\affil[2]{James R. Macdonald Laboratory, Department of Physics, Kansas State University, Manhattan, Kansas, 66506, USA}

\affil[3]{Shanghai Key Lab of Modern Optical System, University of Shanghai for Science and Technology, Shanghai, 200093, China}

\affil[$\ast$]{E-mail: jingsong@ksu.edu}
\affil[$\dagger$]{E-mail: hbjiang@pku.edu.cn}
\affil[$\ddagger$]{E-mail: yunquan.liu@pku.edu.cn}

\begin{abstract}
An intense femtosecond laser pulse can generate ultraviolet air lasing, offering a promising remote light source. A long-standing hypothesis is whether it is seeded by a self-generated spectral component, such as the second harmonic that is inevitably produced by the plasma gradient. Here, we report the generation of both radially and azimuthally polarized N$_2^+$ lasing driven by a single 800-nm cylindrical vector beam. Meanwhile, the same vector pump was applied to drive the generation of vector second harmonics in plasma. The radially polarized pump produces radially polarized second harmonics while the azimuthally polarized pump yields no second harmonic generation owing to the radial direction of plasma gradient. The absence of the azimuthally polarized second harmonic rules out the hypothesis of self-seeding by second harmonics, as both radially and azimuthally polarized N$_2^+$ lasing are observed with comparable intensities. By characterizing the spatial phase distribution of vector 391-nm lasing, we concluded that its phase is synchronized with the pump. These results suggest that amplified spontaneous emissions are the origin of N$_2^+$ lasing under the most common condition of low gas pressure, which was effectively demonstrated by theoretical simulations. Our work provides a promising method for remotely generating vector ultraviolet light sources.
\end{abstract}

\begin{document}

\flushbottom
\maketitle
%
%
\newpage

\section*{Introduction}

Air lasing is a phenomenon in which laser-induced air produces cavity-free coherent light \cite{dogariu2011high,yao2011high,nie2024bidirectional}, effectively acting as a laser. Unlike conventional lasers, which require a physical gain medium (such as a crystal, gas, or liquid) within a cavity to amplify light, air lasing occurs in the atmosphere without the need for a physical cavity. Among various types of air radiations, N$_2^{+}$ air lasing \cite{yao2011high}, particularly the 391-nm radiation corresponding to the transition between ground vibrational states from B to X ionic states, is one of the most debated phenomena. It has garnered significant interest due to its complex and interesting underlying mechanism \cite{xu2015sub,liu2015recollision,zhang2020sub,yao2016population,azarm2017optical,britton2018testing,ando2019rotational,zhang2021coherent,chen2024multiphoton,yuen2024probing,yuen2023coherence,richter2020rotational,gao2024controlling,xie2020role,gao2023structured,hu2023generation,kleine2022electronic,mei2023amplification,shenhigh2024high,lu2025cascaded,cao2025composite,danylo2025measurement}, which results in a long-standing debate on whether there exists population inversion. Most recently, some promising applications of air lasing have been explored, such as coherent Raman spectroscopy\cite{liu2020extremely,zhang2023electronic,lu2024single,zhang2025detection}, ultraviolet supercontinuum light source\cite{lei2022ultraviolet,gao2025controlling,gao2025broadband}, and carrier-envelope phase (CEP) tagging technology\cite{gao2025broadband}. Previous experiments \cite{ni2013identification, xu2018free} have clearly demonstrated that 391-nm lasing can be effectively amplified by an external seed. However, a fundamental question remains unresolved: is the 391-nm lasing process driven by one laser beam seeded by self-generated spectral components, such as self-phase modulation (SPM) or second harmonic generation (SHG)? 

The prevailing use of Ti:sapphire (Ti:Sa) laser pulses complicates this question further, as the wavelength of the driving field is around 800 nm, with its second harmonic overlapping with the 391-nm transition. If the 391-nm lasing is self-seeded, it can be attributed to superfluorescence or light amplification by stimulated emission of radiation (i.e., laser action). Conversely, if the 391-nm lasing process does not involve any seeding, it is strictly amplified spontaneous emission (ASE) \cite{luo2003lasing}. For few-cycle Ti:Sa pulses as the driving field, the spectrum typically spans 550–1000 nm, and the SPM during the focusing process can easily generate a significant supercontinuum extending down to 391 nm at 1 bar pressure \cite{xu2015sub}, serving as a seed. This is reasonable because the intensity and spectral broadening of SPM are proportional to the gas pressure, sample length, and pulse peak intensity \cite{corkum1985generation,perry1994self}. Thus, for the most commonly used multi-cycle Ti:Sa pulses with durations of 35 fs as the driving field, only a very weak supercontinuum was observed near 391 nm in 1 bar \textit{air} with a pulse energy of 6 mJ \cite{chu2013self}. However, within the optimal gas pressure window around 5-50 mbar that yields the strongest and clearest 391-nm lasing, SPM cannot broaden the driving field spectrum down to 391 nm because of the narrow spectral bandwidth (800 nm $\pm$ 14 nm) and low gas pressure. To the best of our knowledge, no direct experimental observation has been reported of a supercontinuum near 391 nm at low pressures pumped by mJ-level 35-fs multi-cycle Ti:Sa lasers. Nevertheless, N$_2^{+}$ lasing at 391 nm can be easily observed from 5 to 30 mbar with a pump pulse energy below 3 mJ \cite{liu2015recollision, chen2019electronic, gao2023structured, hu2023generation, gao2024controlling}, even in a dilute gas jet  \cite{britton2018testing, britton2019short} that is too short to support the sample length needed for significant SPM. This rules out SPM as the self-seeding mechanism of 391-nm lasing in the most common case of 35-fs Ti:Sa pulse at low gas pressure. 

Thus, the prevailing reliable viewpoint is that 391-nm lasing is seeded by the second harmonic generated by the charge gradient in the laser plasma filament \cite{beresna2009high}, which was proposed by one of the present authors \cite{liu2013self}. In contrast to the assumption for SPM as the self-seeding at low pressure, the SHG generated by the charge gradient was indeed experimentally observed at a low pressure of 4 mbar, and it was also experimentally proved by another independent group \cite{hu2023generation}. However, the hypothesis of SHG self-seeding lacks direct and comprehensive evidence and relies only on observing the corresponding spectra, overlooking important spatial information of SH generated from the ponderomotive force acts. Now, we started to consider the spatial information and noticed that the Gaussian-beam-driven air lasing is always the fundamental Gaussian mode (HG$_{00}$ mode), while the Gaussian-beam-driven SHG from the plasma gradient should be HG$_{01}$ or HG$_{01}$ mode that contains two lobes\cite{beresna2009high}. It actually challenges the hypothesis of SHG self-seeding because of the mismatch of the SH mode and the air lasing mode. Therefore, investigating potential self-seeding mechanisms in 391-nm lasing remains an important and unresolved question.

Structured light with a customized profile in space and time can now be well constructed due to the development of laser tailoring techniques \cite{rubinsztein2016roadmap,forbes2021structured,cristiani2022roadmap,shen2023roadmap,fang2025ultrafast}. One dramatic paradigm is cylindrical vector beams (CVBs) that possess cylindrical symmetriclly polarization distribution in space \cite{zhan2009cylindrical}, where the radially polarized and azimuthally polarized cases are shown in Fig. 1(a). A laser-induced plasma filament also exhibits cylindrical symmetry about the light propagation direction, resulting in a charge gradient along the radial direction in the beam cross plane, parallel to the radially polarized electric field and orthogonal to the azimuthally polarized electric field. Therefore, the key idea of our method is to introduce these features of CVBs into the laser-plasma interaction to clarify the role of the SHG in the 391-nm N$_2^+$ lasing.

Here we drive the 391-nm lasing in N$_2$ and the second harmonic in argon with CVBs. We demonstrate the generation of both radially and azimuthally polarized 391-nm N$_2^{+}$ lasing with similar intensities, whereas the second harmonic disappears when switching the driven field from radial to azimuthal polarization. The vector N$_2^{+}$ lasing inherits the same CVB mode as the pump light. In contrast, the radially polarized SH belongs to a new type of radial vector beam with a higher radial order, while the azimuthally polarized pumping yields no SH signal. By analyzing the focal properties of CVBs, we show that the spatial phase of the N$_2^{+}$ lasing is synchronized with the spatial phase of the pump light rather than being double that of the pump light. This indicates that the phase distribution of the driving field is directly transferred to the 391-nm lasing. By combining the results of 391-nm N$_2^{+}$ lasing and the SHG in argon, we conclude that 391-nm N$_2^{+}$ lasing driven by the most commonly used mJ-level multi-cycle Ti:Sa laser pulse at optimal gas window does not involve any self-seeding mechanism and is therefore consistent with ASE.

\section*{Results}
\subsection*{Vector ultraviolet N$_2^+$ lasing}
The experimental setup is depicted in Fig. 1(a) (see Methods for details). An 800-nm CVB (1 kHz, 3 mJ, 35 fs) is employed as the pump light, which is focused on the 10-mbar pure nitrogen by a lens. The N$_2^{+}$ lasing is emitted from the peak density region of the laser-induced plasma. The angle of the q-plate\cite{dorn2003sharper,machavariani2007efficient} is set to make the pump beam radially and azimuthally polarized, respectively, where the fast axis orientation of the q-plate and the spatial polarization distribution of the pump beam are depicted in Fig. 1(a). A linear polarizer is used to characterize the spatial polarization distribution of the N$_2^{+}$ lasing.

\begin{figure}[htbp]
\centering
\includegraphics[width=12.6cm]{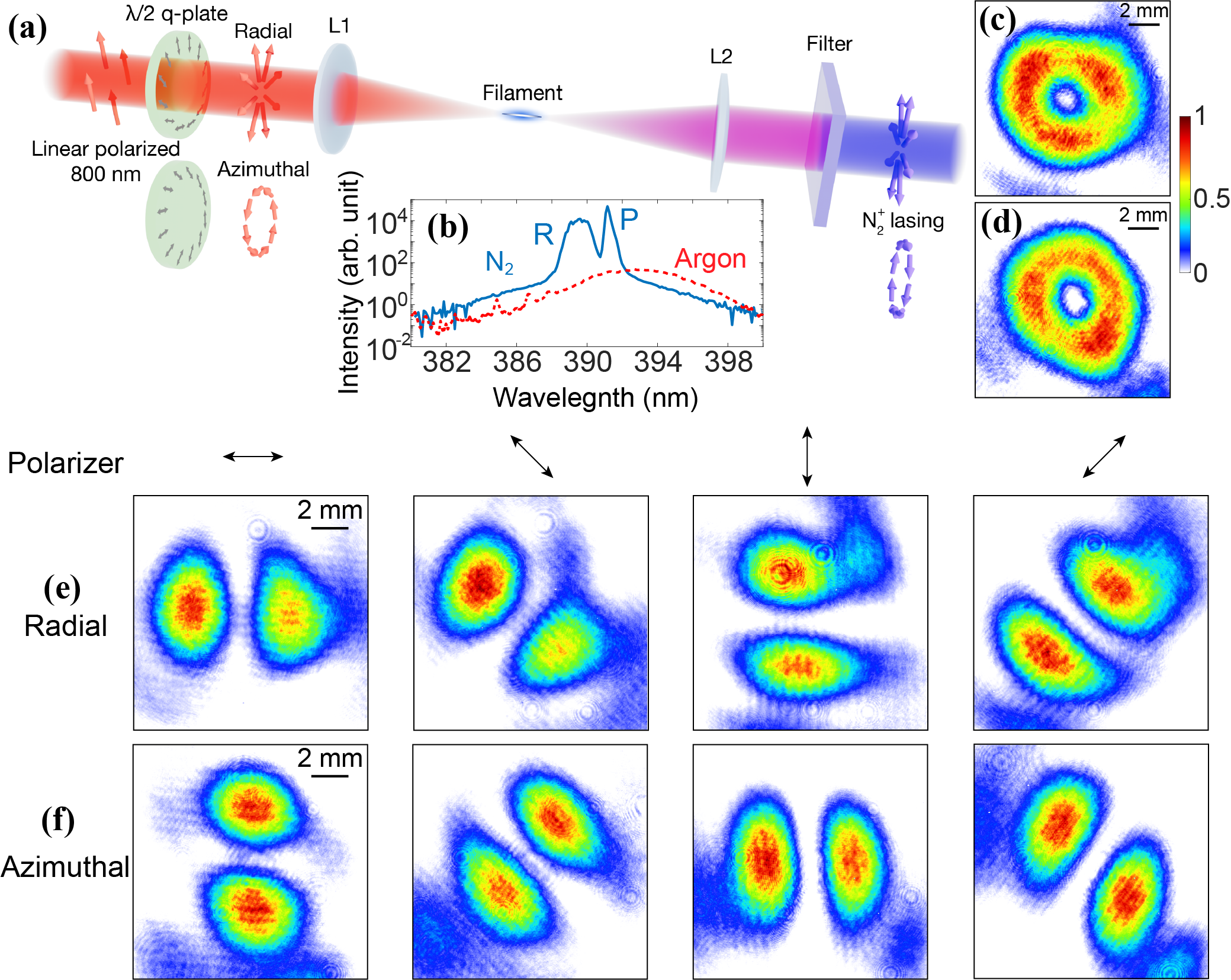}
\caption{\textbf{Generation of vector ultraviolet N$_2^+$ lasing}. (\textbf{a}) Schematic of the experiment. Red and blue arrows represent electric field polarization distributions of CVB pump and vector air lasing, respectively. Grey arrows in q-plate represent fast axis locally. L1, L2 are the lenses with the focal length of 30 cm. The filter is a band-pass filter with a central wavelength of 390 and a bandwidth of 10 nm. (\textbf{b}) Measured 391-nm N$_2^{+}$ lasing spectrum (blue solid line) and the second harmonic spectrum from argon (red dashed line). (\textbf{c, d}) Measured beam profiles of the generated radially (c) and azimuthally (d) polarized 391-nm lasing. (\textbf{e}) Measured radially polarized 391-nm lasing beam profiles after passing a polarizer oriented along different angles marked with black arrows. (\textbf{f}) Corresponding experimental results for azimuthally polarized 391-nm lasing. The color bar applies to all panels c-f with the same normalized standard.}
\label{fig:figure1}
\end{figure}

Figure 1(b) shows measured spectra of the 391-nm N$_2^{+}$ lasing. For the 391-nm lasing, it displays two peaks corresponding to the P and R branches of rotational transitions. The contribution from SPM can be ruled out from the weak and symmetrical pedestals on the two sides of the 391-nm lasing peak. If it is seeded by SPM extended from 800 nm, the right side of the pedestal will show some intensity above the noise level at the wavelength larger than 391 nm \cite{xu2015sub}, and will increase with wavelength. The beam intensity profiles of the generated radially and azimuthally polarized 391-nm lasing are depicted in Fig. 1(c) and Fig. 1(d), respectively. Both profiles exhibit a doughnut-shaped distribution, indicating the presence of a polarization singularity at the beam center. To examine the spatial distribution of the polarization state within the beam plane, we present the beam intensity profiles in Fig. 1(e) and Fig. 1(f) after passing through a polarizer, with the polarizer’s orientation indicated above each figure. Additional polarization analyses, obtained by rotating the polarizer orientation at various positions within the beam profile, are provided in Fig. S4 of Supplementary Note 4 in Supplementary Information. The polarizations at all positions are linear, possessing radial polarization and azimuthal polarization characteristics, respectively. These results confirm that the generated radially and azimuthally polarized 391-nm lasing exhibit approximately equal intensities. Furthermore, the polarization distribution of the 391-nm lasing mirrors that of the driving field in both radial and azimuthal configurations. The laser plasma induced by CVBs features a doughnut-type cross section, similar to a fiber for generating CVBs \cite{ramachandran2009generation}, which enables the generation, propagation, and amplification of vector 391 nm N$_2^+$ lasing due to the cylindrically symmetric gain medium.

\subsection*{Second harmonics generation via plasma driven by cylindrical vector beams}
The SHG can occur in an isotropic gaseous medium if the plasma within the irradiated volume is inhomogeneous. The resulting ponderomotive force acts outward from the center of the pump beam, creating an uneven electron distribution. This redistribution pushes electrons away from the center of the focused beam, ultimately leading to the SHG. The second-order polarization can be written as \cite{beresna2009high,bethune1981optical},
\begin{equation}
    \vec{P}_{2\omega} \propto [1/2 \nabla E_P^2 + \frac{2(\vec{E}_P \cdot \nabla \mathrm{ln}n_e)\vec{E}_P}{\epsilon_p}],
\end{equation}
where $\vec{E}_P$ is the driving electric field in the focal plane, $\epsilon_p$ is the plasma dielectric constant, and $n_e$ is the plasma density. In the nonuniform plasma, the last term in Eq. (1) dominates the SHG. $\nabla \mathrm{ln}n_e$ stands for the electron gradient along the radial direction. Therefore, the product between the driving field and the electron gradient, $\vec{E}_P \cdot \nabla \mathrm{ln}n_e$, will be zero for the azimuthally polarized driving field and be a maximum for the radially polarized driving field.

\begin{figure}[htbp]
\centering
\includegraphics[width=12.6cm]{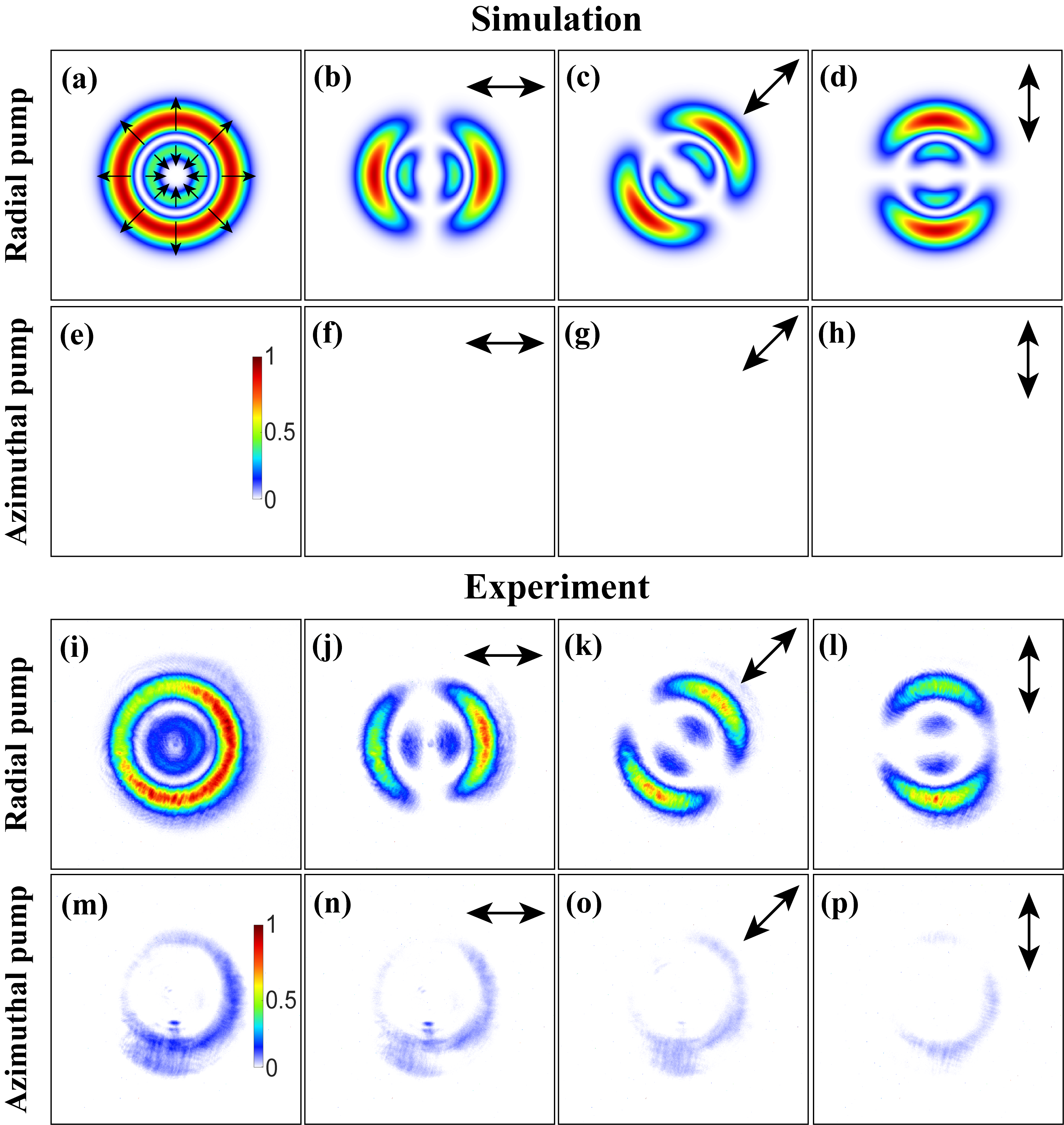}
\caption{\textbf{Second harmonic generation in argon plasma driven by cylindrical vector beams.} The first column shows the beam profile of second harmonics. Columns two through four show the beam profile after passing a polarizer with the orientation marked in the upper-right corner. (\textbf{a-d}) Calculated beam intensity profile driven by radially polarized light fields. The arrows in (a) indicate the polarization state locally. (\textbf{e-h}) Calculated beam intensity profile driven by azimuthally polarized light fields. The intensity is rigorously zero in Eq. (1) for the case of azimuthal polarization. The intensities in the simulation share the color bar with the same standard in (e). (\textbf{i-l}) Measured beam intensity profile driven by radially polarized light fields. (\textbf{m-p}) Measured beam intensity profile driven by azimuthally polarized light fields. The intensities in (m-p) are normalized to those in (i-l).}
\label{fig:figure2}
\end{figure}

To eliminate the strong disturbance from N$_2^+$ lasing and to observe a clean beam profile of a pure SHG signal on the CCD, we select argon as the target gas in the experiment, since it exhibits no significant atomic emission lines around 400 nm. Changing gas target for comparison is reasonable because SHG beam pattern is independent of gas species, and is determined only by the charge-density gradient. Moreover, argon and N$_2$ have very close ionization potentials (15.76eV and 15.58eV, respectively), which leads to nearly identical charge densities under the same laser condition. The measured SH spectrum is shown in Fig. 1(b), which is close to the wavelength of N$_2^+$ lasing. As discussed above, the SH emission is generated at the peak of the plasma density gradient. In Fig. 2, we illustrate the theoretical and experimental results of the SH driven by CVBs in argon. Fig. 2(a) shows the calculated SH beam profile intensity with the arrows indicating the polarization state locally, driven by the radially polarized light field, based on Eq. (1). In Figs. 2(b-d), we show the corresponding beam profiles after passing a polarizer oriented along horizontal, 45$^\circ$ and vertical directions, respectively. The numerical results are obtained by solving wave equation, that is, Eq. (2), where the simulation details are in Methods). The theoretical result shows there are two concentric rings in the beam profile with a phase difference of $\pi$, which originates from the gradient of the doughnut-type plasma having two opposite peaks in the radial direction. This is a new type of CVBs similar to a radially polarized Bessel-Gaussian beam \cite{schimpf2013radially,vyas2014generation}, but with only two concentric rings. It can be approximately described as the superposition of a $x$-polarized HG$_{30}$ mode and a $y$-polarized HG$_{03}$ mode, whereas the driving field and the corresponding N$_2^+$ lasing are both the superposition of a $x$-polarized HG$_{10}$ mode and a $y$-polarized HG$_{01}$ mode \cite{zhan2009cylindrical}. In contrast, in the azimuthally polarized case, the gradient of the laser plasma is perpendicular to the driving electric field, i.e., $\vec{E}_P \cdot \nabla \mathrm{ln}n_e=0$. Thus, the calculated SH intensity must be zero. For intuitive comparison, we also show the calculation results in Figs. 2(e-h). We have theoretically demonstrated that radially polarized driving fields can generate SH, whereas azimuthally polarized driving fields cannot generate SH at all.

We proceed to experimentally verify the calculation results. The experimental measurement results of the SHG driven by radially and azimuthally polarized light fields are shown in Figs. 2(i-l) and Figs. 2(m-p), respectively. The result in the radial case indeed shows two rings in the beam profile, matching the numerical results shown in Figs. 2(a-d), and presents a completely different CVB mode compared with the radially polarized N$_2^+$ air lasing in Figs. 1(c,e). More importantly, the measured SH intensity by azimuthally polarized fields is near zero, also agreeing well with the simulation in  Figs. 2(e-h). Notably, the weak signal appears in Fig. 2(m) because the cylindrical symmetry of the driving field cannot be ideal in the experiment. Even though there is a weak SH signal, its azimuthal polarization information is totally lost, as shown in Figs. 2(n-p). With excellent agreement with simulations, our experimental results convincingly proved that radially polarized driving fields can produce SHG, whereas azimuthally polarized driving fields cannot.

 The absence of SHG in the azimuthal case indicates that SH plays a minimal role in the 391-nm lasing, as we have demonstrated the generation of vector N$_2^+$ lasing for both radial and azimuthal polarizations with approximately equal intensities. This is reasonable and consistent with the theoretical expectation because the effective gain region is located near the peak of the driving field’s intensity, where the gradient of the plasma is almost zero, i.e., there is no SHG. As a result, in the laser filamentation, the SH and the gain region are inherently not overlapped in space. 

 On the other hand, if SH accounts for self-seeding, the intensity of radially polarized N$_2^+$ lasing in Fig. 1(c) should be expected to be significantly stronger than that of azimuthally polarized N$_2^+$ lasing in Fig. 1(d), because of the existence of the radially polarized SH and the absence of the azimuthally polarized SH shown in Fig. 2. Furthermore, if N$_2^+$ lasing is self-seeded by SH, the high-order pattern of radially polarized SH shown in Figs. 2(a-d,i-l) will be maintained in the amplification of the radially anisotropic gain medium \cite{gao2023structured} that was induced by the permanent alignment of N$_2^+$ \cite{gao2024controlling,xie2024anisotropic}. This can be readily verified by the model we previously proposed \cite{gao2023structured}. However, the pattern of the radially polarized SH is different from that of the radially polarized N$_2^+$ lasing. The mismatch between air lasing and SHG in both intensity and spatial mode strongly rules out the self-seeding role of SHG.

\subsection*{Spatial phase measurement}
Further experimental evidence from the perspective of spatial phase is provided in Figure 3. The key feature of the seeded air lasing is that the amplified light retains the phase of the seed \cite{mei2023amplification,shenhigh2024high}, which can be a criterion for checking the 391-nm lasing driven by a single laser beam. In Fig. 3, we used a cylindrical lens to examine the spatial phase distribution in the beam profile. If the 391-nm lasing is phase-synchronized with the driving field, as depicted in Fig. 3(a), the two lobes after passing through a polarizer should exhibit opposite phases. Consequently, using a cylindrical lens, the two lobes cannot be focused together, and the resulting intensity pattern in the focal plane would still show two distinct stripes, as shown in Fig. 3(c). Conversely, if the 391-nm lasing is seeded by SH, the spatial phase would double. In this scenario, shown in Fig. 3(b), the two lobes after passing through a polarizer would have the same phase, allowing them to focus onto a single strip, as shown in Fig. 3(d). This would produce a dominant beam in the focal plane. Here, we intentionally assume the case of the phase doubling (actually, the SH from the plasma gradient is distinct from a normal parametric process-induced SH). In Methods, we present simulation details for the calculation of the beam focusing with the modified Richard-Wolf vectorial diffraction method for a cylindrical lens, which is described by Eqs. (3-4). Figure 3(e) illustrates our experimental results, which align with the calculated case for phase synchronization, confirming the phase relationship of the 391-nm lasing with the driving field. Notably, we only illustrate the case of the y component of the radially polarized 391 nm lasing here. Indeed, the measured phase differences of the two lobes in all panels of Figs. 1(e-f) are $\pi$.

\begin{figure}[htbp]
\centering
\includegraphics[width=12.6cm]{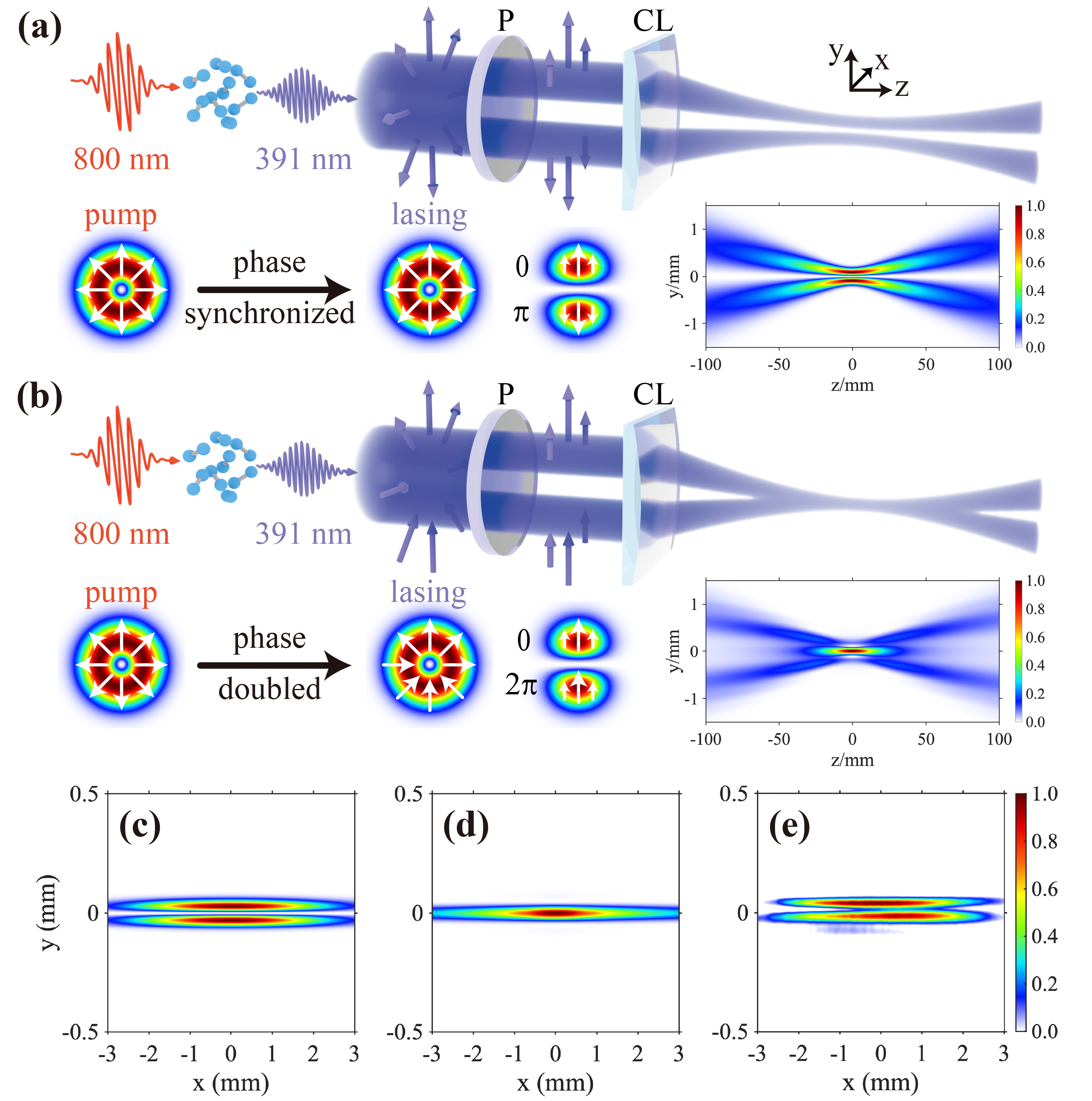}
\caption{\textbf{Phase distribution measurement of the radially polarized 391-nm N$_2^{+}$ lasing by a cylindrical lens.} (\textbf{a}) Calculated results when the spatial phase of 391-nm lasing is synchronized with the driving field. After passing through a polarizer, the two lobes have a phase difference of $\pi$. Blue and white arrows represent local polarization direction of air lasing. P is polarizer, CL is cylindrical lens. (\textbf{b}) Calculated results when the spatial phase of 391-nm lasing is double that of the driving field. After passing through a polarizer, the two lobes have a phase difference of $2\pi$. Blue and white arrows represent local polarization direction of air lasing. P is polarizer, CL is cylindrical lens. (\textbf{c, d}) Calculated beam profiles in the focus plane in the scenarios of (a) and (b), respectively. (\textbf{e}) Experimental result.}
\label{fig:figure3}
\end{figure}

It is worth mentioning that the N$_2^+$ air lasing is a non-parametric process and thereby is fundamentally different from high harmonic generation \cite{gariepy2014creating} and other normal parametric processes \cite{hancock2021second} in which the phase of \textit{n}th harmonic is \textit{n} times that of the driving field. Even in a parametric process such as SPM, any wavelength within the SPM-induced white light supercontinuum remains phase-synchronized with the driving field \cite{tokizane2009supercontinuum,xu2022powerful}. Therefore, the fact that 391 nm is nearly twice the frequency of 800 nm does not imply that the phase of the air lasing must be twice that of the driving field.

\section*{Discussion}

A supercontinuum by SPM can be characterized by B-integral \cite{perry1994self}, namely, $B=2\pi/\lambda\int_0^zn_2|E(z)|^2dz$, where $n_2$ is the nonlinear refractive index linearly proportional to the gas pressure \cite{bree2010method}. It indicates that the ability of spectral broadening is generally proportional to the gas pressure, propagation distance, and peak intensity of light (pulse energy over pulse duration) \cite{corkum1985generation}. In previous experiments, the SPM extending spectrum from 800 nm to 391 nm was experimentally observed only in air at 1 bar; it is relatively strong with a broadband few-cycle pump \cite{xu2015sub} but very weak with a conventional multi-cycle pump with pulse energies of 6 mJ \cite{chu2013self} and 8 mJ \cite{wang2013forward}, whose pulse energies are relatively high. Nevertheless, for most common cases of N$_2^+$ air lasing studies, 391-nm lasing was driven by a multi-cycle 35-fs Ti:Sa laser under the condition of an optimal gas pressure window below 100 mbar \cite{liu2013self}, where 391-nm lasing can be easily observed from 5 to 30 mbar with a pump pulse energy below 3 mJ \cite{liu2015recollision, chen2019electronic, gao2023structured, hu2023generation, gao2024controlling}. Moreover, pulse energy and gas pressure can even reach 1.1 mJ and 5 mbar \cite{liu2017unexpected}, and propagation distance can tolerate the length of a dilute gas jet \cite{britton2018testing, britton2019short}. Obviously, the B-integrals in these common cases can be about 2 or 3 orders of magnitude smaller than that of the 1-bar case. However, just a twofold decrease in the B-integral, below an order-of-magnitude reduction, can markedly suppress spectral broadening \cite{tsai2022nonlinear}. Consequently, there is no previous experimental evidence reported that a multi-cycle 35-fs Ti:Sa laser can extend its narrow-band spectrum from 800 nm to 400 nm under low pressure. In our experiment, even a side fluorescence signal can be easily detected over a small solid angle \cite{gao2024controlling}, whereas the supercontinuum with high collimation was never observed around 400 nm despite capturing the whole forward emission cone. Even in our recent experiment employing 3.7-fs single-cycle pulses as the driving field, it remains difficult to extend the super-broad spectrum of the driving field down to 400 nm at low pressure \cite{gao2025controlling}. Therefore, SPM should not be treated as the self-seeding for the most common case driven by multi-cycle laser pulses at low gas pressure. Citing the white light from SPM obtained at 1-bar gas pressure as evidence of the self-seeding mechanism for the most common low-pressure case is not appropriate.

On the other hand, as shown in Figs. 1 and 2, we can generate both radially and azimuthally polarized air lasing with comparable intensities; however, because of a well-understood physical mechanism, SHG from the plasma gradient occurs only for the radial case, not for the azimuthal case. As we discussed above, due to the existence of air lasing and the absence of SH in azimuthally polarized pumping, it is solid evidence that SH contributes a minimal part to the 391-nm lasing. Therefore, after ruling out the SPM and SHG as the self-seeding, we believe the 391-nm lasing originates from the ASE process. Now, we will discuss how ASE becomes a vector beam in N$_2^+$ lasing.

It is generally accepted that the wavefronts of ASE from different individual emitters are incoherent with each other. Although the phase of each wavefront is uniform, the ensemble of all emitters manifests as unpolarized fluorescence. However, previous studies point out that the spatial coherence of ASE is determined by gain geometry \cite{redding2011spatial,tommasini2000coherence,bachelard2011wavefront}. With a specific anisotropic gain medium, spontaneous emissions could be amplified into a highly linearly polarized beam \cite{ahn2023electrically}, an optical vortex \cite{ma2021vortex}, a CVB \cite{li2023monolithically}, and a spatially coherent light source via Dicke superradiance and subradiance \cite{gold2022spatial}. In our previous study \cite{gao2024controlling}, based on the MO‐ADK theory \cite{tong2002theory,pavivcic2007direct}, we have demonstrated—both experimentally and theoretically—that preferential ionization of N$_2$ induces a permanent alignment of N$_2^+$. This permanent alignment leads to an anisotropic gain with a maximum amplification direction along the pump light's polarization, which is independently proved by another theoretical work with the perspective of the anisotropic quantum coherence \cite{xie2024anisotropic}. Thus, if the pump light is a CVB, the gain will exhibit cylindrical symmetry, the theoretical model of which was built in our previous study \cite{gao2023structured}. By the way, previous studies have demonstrated that transient alignment can modulate air lasing intensity under a pump-probe scheme \cite{zhang2013rotational,xu2017alignment}. This is because, combining permanent alignment, the averaged alignment degree changes with time delay. However, the existence of permanent alignment makes the time-averaged alignment degree always greater than 1/3, making air lasing polarization align with that of the pump light. Here, due to the absence of the seed (single pump), we only employed the theoretical model for the simulations about how ASE becomes a vector beam, where the effect of transient alignment is ignored. The details of the model and simulations are in Methods, which is described by Eqs. (5-6).

In Fig. 4, we show simulations of ASE processes in the radially anisotropic gain when the driving electric field is a radially polarized beam. Following the principle of ASE, the seed is a random spontaneous-emission source including some random noise light spots, where each light spot has a random amplitude, a random phase, a random polarization, and a random location, as shown in the first column of Fig. 4. The energies of these seeds are photon-level. Figs. 4(a-d) show the amplification and propagation processes of ASE wavefronts originating from four individual emitters of random spontaneous emissions, respectively. The amplification factors from photon level in the first column to nanojoule in the fifth column are about $10^{10}$ times. As the simulations demonstrate, regardless of the initial distribution of spontaneous-emission noise, the output ASE wavefront is always radially polarized. This is because, as Eqs. (5-6) present, $E_x$ and $E_y$ are coupled with each other in CVB-induced cylindrically symmetric gain by the macroscopic polarization of N$_2^+$ medium. As a result, even though the individual light spots in the initial spontaneous-emission seed are mutually incoherent, the competition during amplification and propagation through the gain medium synchronizes spatial phase, yielding a coherent, radially polarized ASE wavefront. Here, we show only four cases; in fact, we performed many independent runs starting with different random spontaneous emissions, and in every case the ASE ultimately evolved to be radially polarized. Therefore, the ensemble of all ASE wavefronts collectively manifests as a radially polarized beam, the same as experimental results presented in Figs. 1(c,e).

\begin{figure}[htbp]
\centering
\includegraphics[width=12.6cm]{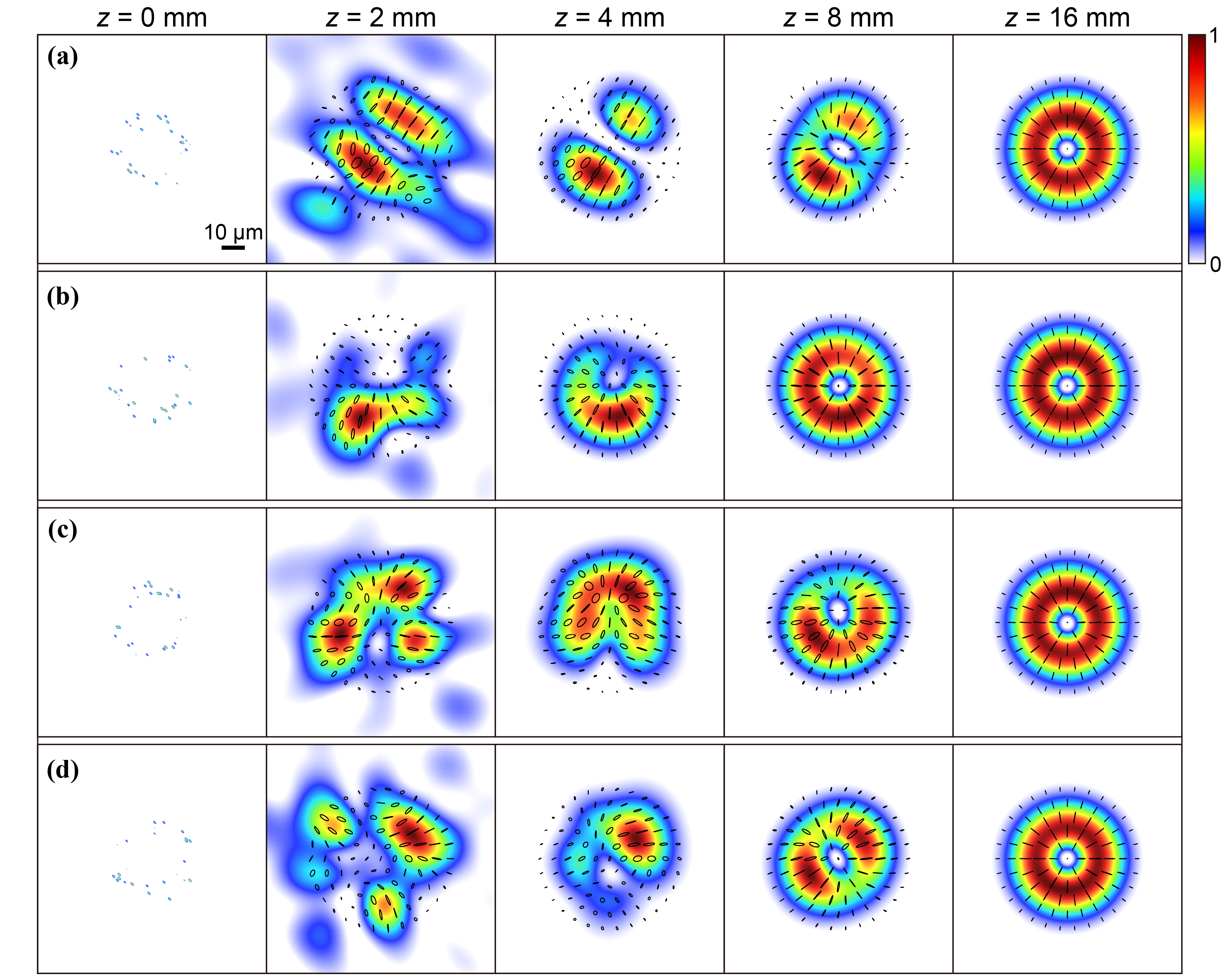}
\caption{\textbf{Simulated amplified spontaneous emissions in a radially anisotropic gain medium.} (\textbf{a-d}) Beam profiles of ASE at different propagation distances in the radial gain, each amplified from an independent random spontaneous-emission source shown in the first column. Each spontaneous-emission source has some random noise light spots with random amplitudes, random phases, and random locations. Columns two through five correspond to propagation distances of 2 mm, 4 mm, 8 mm, and 16 mm, respectively. Black ellipses and lines represent directions and states of polarization locally. A shared color bar is used for all panels.}
\label{fig:figure3}
\end{figure}

In Fig. 5, we show simulations of ASE processes in the azimuthally anisotropic gain when the driving electric field is an azimuthally polarized beam. Similarly, regardless of the initial distribution of spontaneous emission noise, every output ASE wavefront consistently evolved into an azimuthally polarized state, collectively manifesting as an azimuthally polarized beam. This is consistent with the experimental result shown in Figs. 1(d,f). In addition, when the driving electric field is an arbitrary CVB between radial and azimuthal polarization states, all of the ASE wavefronts will exhibit the same vectorial mode as the driving field. Furthermore, when the driving electric field is linearly polarized, the ASE wavefronts will be linearly polarized with the same polarization direction as the driving field. Conversely, when there is no permanent alignment of N$_2^+$, i.e., $<\cos^2{\theta}>$ equals to $<\sin^2{\theta}>$ in the theoretical model, the gain medium will be isotropic. As the corresponding simulations presented in Fig. S3 in Supplementary Note 3 of Supplementary Information, the individual output ASE wavefronts amplified in the isotropic gain medium exhibit random polarization states and thereby are incoherent with each other, collectively manifesting as unpolarized fluorescence. This is consistent with normal cases of ASE \cite{siegman1986lasers,redding2011spatial,li2002time,bachelard2011wavefront,tommasini2000coherence}.

Generally, in a gain medium with a fixed cross-section and a fixed gain coefficient, the intensities of forward and backward ASEs are comparable. However, the laser filament is tapered \cite{hu2023generation}, consisting of a long narrow-to-broad section followed by a short broad-to-narrow section. In such a narrow-to-broad geometry, the backward ASE experiences a gradually decreasing gain cross-section along the propagation direction, which makes it more prone to diffract out of the gain region than the forward case. Previous study demonstrated that forward emissions are more than three orders of magnitude stronger than backward emissions in this tapered gain medium \cite{kerttula2012tapered}. More importantly, the gain lifetime with significant amplification in N$_2^+$ lasing is about 10-20 ps \cite{zhang2013rotational,lei2017population,miao2021optical}, which corresponds to 3-6 mm optical path. This makes the gain coefficient fundamentally different in the forward emission and backward emission: the forward emission experiences a fixed gain coefficient because the time between ions excited by the pump light and stimulated by the forward emission is fixed; in contrast, the backward emission experiences a continuously decreasing gain coefficient because the time delay between exciting and stimulating of ions increases with propagating, resulting in almost no optical amplification at the second half of the backward propagation in laser plasma. These two aspects make backward ASE hard to be detected in the experiment.

To be summarized, as previous studies present \cite{ahn2023electrically,ma2021vortex,li2023monolithically}, not all cases of ASE manifest as unpolarized fluorescence. In our case, we demonstrated that, owing to the permanent alignment of N$_2^+$, the amplification of randomly spontaneous-emission seeds at 391 nm into CVBs is an inevitable outcome in such a cylindrically symmetric anisotropic gain medium.

\begin{figure}[htbp]
\centering
\includegraphics[width=12.6cm]{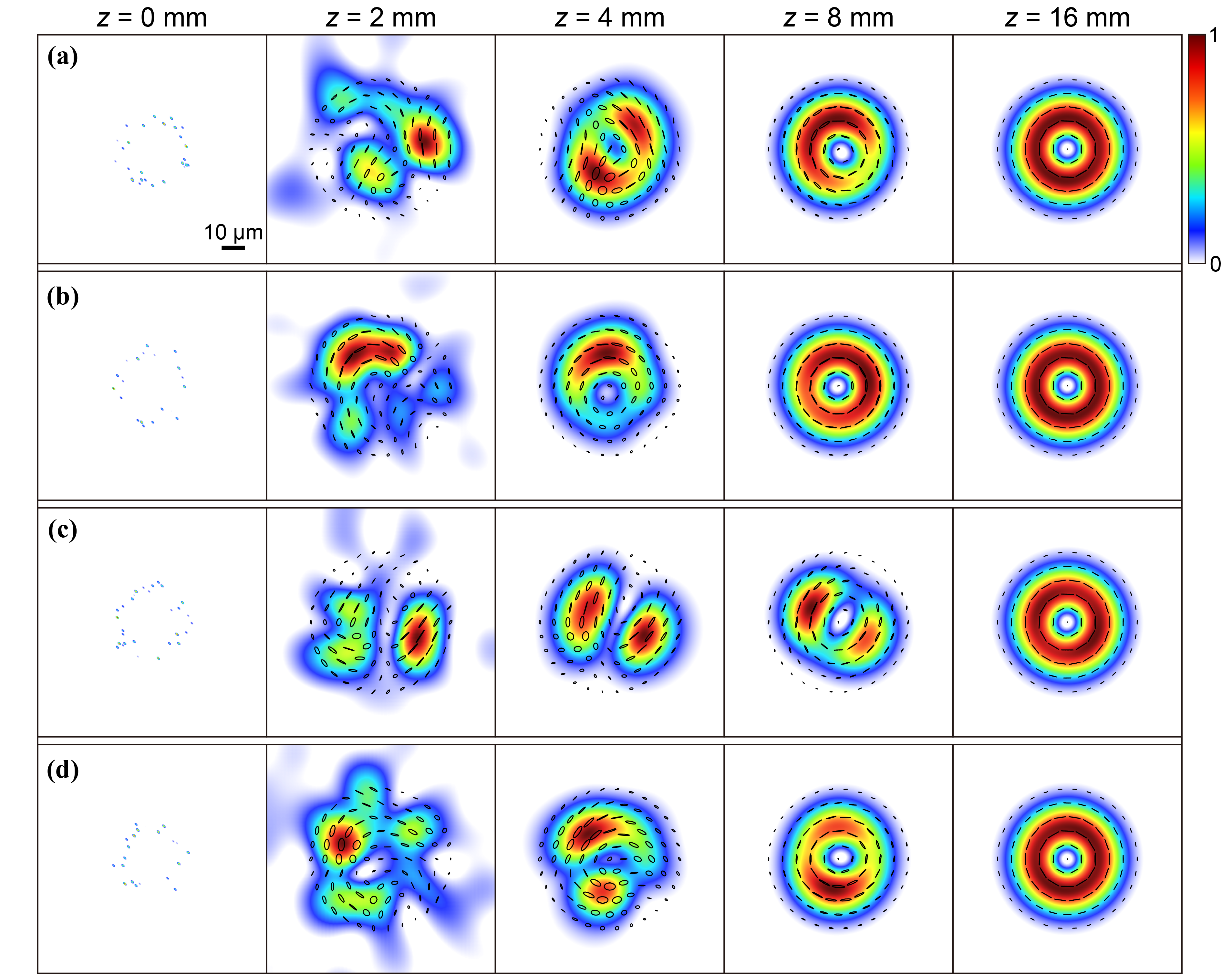}
\caption{\textbf{Simulated amplified spontaneous emissions in an azimuthally anisotropic gain medium.} (\textbf{a-d}) Beam profiles of ASE at different propagation distances in the azimuthal gain, each amplified from an independent random spontaneous-emission source shown in the first column. Each spontaneous-emission source has some random noise light spots with random amplitudes, random phases, and random locations. Columns two through five correspond to propagation distances of 2 mm, 4 mm, 8 mm, and 16 mm, respectively. Black ellipses and lines represent directions and states of polarization locally. A shared color bar is used for all panels.}
\label{fig:figure3}
\end{figure}

\section*{Conclusions}

In conclusion, our study provides critical insights into the generation mechanism of N$_2^{+}$ lasing at 391 nm and the ponderomotive force acts in plasma, specifically addressing the role of self-seeding by the second harmonic. We generated both radially and azimuthally polarized 391-nm lasing driven by a single pump beam. By analyzing the spatial phase distribution, we demonstrated that the lasing phase is synchronized with the driving field. Because azimuthally polarized pumping yields no SH signal yet still produces azimuthally polarized air lasing with an intensity comparable to that of the radially polarized case, second harmonic self-seeding is effectively ruled out. Otherwise, the intensity of radially polarized air lasing should be expected to be significantly stronger than that of the azimuthally polarized one. Our experimental results suggest the 391-nm lasing originates from the ASE process under the most common condition of low pressure and 800-nm multi-cycle pulse pumping. It should be mentioned that, compared with ASE, SPM can be a significant self seed under the condition of relatively high pulse energy at 1-bar gas pressure or few-cycle pulse pumping, owing to the related strong spectral broadening ability. We then performed simulations to explain how ASE evolves into a CVB with a theoretical model of cylindrically symmetric gain medium that is derived from the experimentally confirmed permanent alignment of N$_2^+$. We also demonstrated the implementation of a special high-order radially polarized SH from the gradient of the laser plasma. This work not only resolves a key aspect of the lasing mechanism but also introduces a method for remotely generating ultraviolet vectorially structured light fields, offering new possibilities for applications in advanced remote sensing and laser physics.

\section*{Methods}
\subsection*{Experimental Setup}
The s-polarized Gaussian laser pulses are launched from a Ti:Sa amplifier (Spectra-Physics Spitfire Ace) with a central wavelength of 800 nm, a pulse duration of 35 fs, a repetition of 1 kHz, and a single pulse energy of 3 mJ. After passing through a thin half-wave q-plate, the 800 nm Gaussian beam becomes a CVB with spatially variant polarization. Note that the q-plate was selected intentionally to avoid generating SH by itself. Then, the CVB is focused on the 10-mbar pure nitrogen or argon gas by a lens with a focal length of 30 cm. The corresponding peak intensity at the focus is estimated to be the order of $10^{14}$ W/cm$^{2}$ regarding defocusing effect and relatively low peak intensities of CVBs. In nitrogen, the N$_2^{+}$ lasing is emitted from the peak density region of the laser-induced plasma. A 650-nm short-pass filter was used to move out the driving field. Then, we directly measure the spectrum by a fiber head connected to a grating spectrometer. In order to capture a pure beam profile of N$_2^+$ lasing, a band-pass filter with a central wavelength of 390 nm and a bandwidth of 10 nm was used to move out other spectral components. After the band-pass filter, the beam profile was recorded by a CCD. In argon, the SH emission, generated at the peak of the plasma density gradient, is recorded by the CCD and the spectrometer after the 650-nm shortpass filter. The angle of the q-plate is set to make the pump beam radially and azimuthally polarized, respectively, where the fast axis orientation of the q-plate and the spatial polarization distribution of the pump beam are depicted in Fig. 1(a). A linear polarizer is used to characterize the spatial polarization distribution of the N$_2^{+}$ lasing and the SH. A cylindrical lens with a focal length of 1 m is used to measure the spatial phase of vector air lasing. 

\subsection*{Numerical Simulations}
\noindent\textbf{Second harmonic generation simulations}. The polarization for second harmonic generation (SHG) from the ponderomotive force in plasma
is given by Eq. (1). In laser-induced plasma, the second term on the right side dominates the SHG. This is because the first term is a purely irrotational polarization without curl, giving rise to some longitudinal mode and thereby no radiation at far field\cite{bethune1981optical}. Thus, we neglect the first term and substitute Eq. (1) in the wave equation under the condition of a slowly varying approximation, namely,
\begin{equation}
   \frac{\partial^2 \vec{E}_{SH}}{\partial x^2}+\frac{\partial^2 \vec{E}_{SH}}{\partial y^2}+2ik\frac{\partial\vec{E}_{SH}}{\partial z}=-\frac{2\mu\omega^2\chi^{(2)}}{\epsilon_p}(\vec{E}_P \cdot \nabla \mathrm{ln}n_e)\vec{E}_P,
\end{equation}
where $\vec{E}_{SH}$ is the electric field of SHG, $k$ is the wavenumber of the driving electric field, $\omega$ is the angular frequency of the driving electric field, and $\mu$ is the permeability. The equations are solved using the split-step Fourier method. In the calculation for the driving electric field of a CVB, we neglected the z-direction dependence of the plasma density because we only considered the spatial transverse distribution of the electric field of the SHG (see Supplementary Note 1 of Supplementary Information for numerical details).

\noindent\textbf{Focusing simulations}.We applied the principle of Richard-Wolf vectorial diffraction method\cite{richards1959electromagnetic} to simulate the focusing process of a CVB. In the experiment, we used a cylindrical lens instead of a spherical lens to focus the vector air lasing. A cylindrical lens focuses the light in only one dimension (x or y direction), allowing us to observe two clear and distinct stripes rather than two barely noticeable and tiny spots. In order to make the simulation consistent with the experimental setup, we modified the Richard-Wolf method so that it can be applied to a cylindrical lens (See supplementary Note 2 of Supplementary Information). Thus, the electrical field after a cylindrical lens without the approximation near the focus is given by, 
\begin{equation}
    \vec{E}(x,y,z) =-\frac{ik}{2\pi}\int_{-\theta'_m}^{\theta'_m}\int_{-\frac{D}{2}}^{\frac{D}{2}}P(\theta')E(x',y')\vec{u}\frac{e^{ikR}}{R}K(\chi)dS,
\end{equation}
 where $\vec{E}(x,y,z)$ is the electric field after a cylindrical lens, $(x,y,z)$ is the corresponding Cartesian coordinate system, $k$ is the wavenumber, $(x',y',z')$ is the Cartesian coordinate system with the center of the lens as the origin before focusing, $\theta'$ is the polar angle in the output pupil of the focusing system given by $y'=f\sin{\theta'}$, $\theta'_m$ is the maximum angle determined by the numerical aperture of the lens ($\frac{D}{2}=f\sin{\theta'_m}$, $f$ is the focal length, $D$ is the diameter of the aperture), $P(\theta')=\sqrt{\cos{\theta'}}$ is the pupil apodization function, $E(x',y')$ is the amplitude of incident electric field before the lens in the pupil plane, $z'$ is parallel to $z$, $K(\chi)=\cos{\chi}$ is the inclination factor, and $\chi$ is the angle between $\vec{R}$ and $\vec{s}$. As shown in Fig. S2 in Supplementary Note 2 of Supplementary Information, $\vec{s}$ is the unit vector of propagation direction of the ray after the lens, $\vec{r}$ is the position vector with the focus as the origin, and $\vec{R}$ indicates the displacement from the pupil plane to the image plane. Noticing that $P(\theta)E(x',y')\vec{u}$ represents the electrical field after the lens in the pupil plane, where $\vec{u}$ is given by
\begin{equation}
    \vec{u}=\cos{\phi}
    \begin{pmatrix}
        \frac{x'}{\sqrt{x'^2+y'^2}}\vec{e}_x \\
        \frac{y'}{\sqrt{x'^2+y'^2}}\cos{\theta'}\vec{e}_y\\
        \frac{y'}{\sqrt{x'^2+y'^2}}\sin{\theta'}\vec{e}_z
    \end{pmatrix}
    +\sin{\phi}
    \begin{pmatrix}
        -\frac{y'}{\sqrt{x'^2+y'^2}}\vec{e}_x \\
        \frac{x'}{\sqrt{x'^2+y'^2}}\cos{\theta'}\vec{e}_y\\
        \frac{x'}{\sqrt{x'^2+y'^2}}\sin{\theta'}\vec{e}_z
    \end{pmatrix},
\end{equation}
where $\phi$ is a factor determining the percent of radial and azimuthal amplitude, $\vec{e}_x$, $\vec{e}_y$, and $\vec{e}_z$ are the unit vectors of x, y, and z directions, respectively. It should be noted that $\phi=0^\circ$ represents radial polarization and $\phi=90^\circ$ represents azimuthal polarization. In the case of a cylindrical lens, variables follow the relationships that are $\vec{s}=-\sin{\theta'}\vec{e}_y+\cos{\theta'}\vec{e}_z$, $\vec{r}=x\vec{e}_x+y\vec{e}_y+z\vec{e}_z$, $\vec{x'}=x'\vec{e}_x$, $\vec{R}=f\vec{s}-\vec{x'}+\vec{r}$, and $dS=dx'dy'=fdx'd\theta'$. Combining Eq. (3) and Eq. (4), we can simulate x, y and z component of electric field distribution at any distance in focusing process (See supplementary Note 2 of Supplementary
Information for numerical details).

\noindent\textbf{Amplified spontaneous emission simulations}. In our previous study \cite{gao2024controlling}, based on the MO‐ADK theory \cite{tong2002theory,pavivcic2007direct}, we have demonstrated that preferential ionization of N$_2$ induces a permanent alignment of N$_2^+$. We show that the permanent alignment leads to an anisotropic gain with a maximum amplification direction along the pump light's polarization, which is also proved by another theoretical work with the perspective of the anisotropic quantum coherence \cite{xie2024anisotropic}. As a result, when the pump light is a cylindrical vector beam, the corresponding anisotropic gain medium will exhibit cylindrical symmetry. We derived the theory model of this cylindrically anisotropic gain medium in our previous work, where the corresponding macroscopic polarization is given by \cite{gao2023structured}
\begin{equation}
    P_x = C[<\cos^2{\theta}>(E_x\cos^2{\varphi}+E_y\sin{\varphi}\cos{\varphi})+<\sin^2{\theta}>(E_x\sin^2{\varphi}-E_y\sin{\varphi}\cos{\varphi})],
\end{equation}
and
\begin{equation}
    P_y = C[<\cos^2{\theta}>(E_x\sin{\varphi}\cos{\varphi}+E_y\sin^2{\varphi})-<\sin^2{\theta}>(E_x\sin{\varphi}\cos{\varphi}-E_y\cos^2{\varphi})],
\end{equation}
where $P_x$ and $P_y$ are the x and y components of the macroscopic polarization, respectively, $C$ is a constant, $\varphi$ is the azimuthal angle in $xy$ plane, $E_x$ and $E_y$ are the x and y components of the electric field of 391 nm lasing, respectively, $<\cos^2{\theta}>$ and $<\sin^2{\theta}>$ are the permanent alignment degrees of $N_2^+$ that are parallel to the radial and azimuthal directions, respectively. As we can see, in the case of a cylindrical vector pump beam, the macroscopic polarization is a function of both the x and y components of the seed electric field, namely, $E_x$ and $E_y$ are coupled with each other in such a gain medium. Substituting Eq. (5) and Eq. (6) into wave equation, we can simulate how ASE evolve into a CVB during process of propagation and amplification, as shown in Fig. 4 and Fig. 5 (see Supplementary Note 3 of Supplementary Information for numerical details).

\section*{Data availability} 
The data that support the findings of this study are available from the
corresponding author on request.

\section*{Code availability} 
The code that supports the findings of this study is available from the
corresponding author on request.

\bibliography{ref}

\section*{Author contributions}
J. G., H. J. and Yunquan L. conceived the study. J. G. performed the experiments with assistance from Y. W. and H. M.. J. G. and H. J. built the cylindrical symmetric anisotropic gain model. J. G. extended the Richard-Wolf vectorial diffraction method to the cylindrical lens case. J. G. performed the simulations. J. G., Yi L., C. W., H. J., M. H., and Yunquan L.  analyzed and interpreted the data. C. W. and Yunquan L. provided elements of the experimental setup. J. G. and J. D. designed all schematics. M. H. and J. G. drafted the paper with the input from all authors.

\section*{Acknowledgments}
This work was supported by the National Key R\&D Program (No. 2022YFA1604301) and the Natural Science Foundation of China (No. 12334013, No. 92250306, No. 12595343, and No. 12404393).


\section*{Competing interests} 
The authors declare no competing interests

\clearpage

\end{document}